\documentclass[twocolumn,traditabstract,longauth,letter]{aa}
\usepackage{amssymb}
\usepackage{mathptmx}
\usepackage{natbib}
\usepackage[]{graphicx}
\defcitealias{canameras15}{C15}
\begin{document}

\title{Planck's Dusty GEMS. II. Extended [CII] emission and absorption
  in the Garnet at z$=$3.4 seen with ALMA\thanks{Based on data
    obtained with ALMA in program 2013.1.01230.S, and with EMIR on the
    IRAM 30~m telescope in program 223-13.}} 
\author{N.~Nesvadba\inst{1,2,3,4}~\thanks{nicole.nesvadba@ias.u-psud.fr},
  R.~Kneissl\inst{5,6}, R.~Canameras\inst{1,2,3,4}, F.~Boone\inst{7},
  E.~Falgarone\inst{8}, B.~Frye\inst{9}, M.~Gerin\inst{8},
  S.~Koenig\inst{10}, G.~Lagache\inst{11}, E.~Le Floc'h\inst{12}, 
  S.~Malhotra\inst{13},
  D.~Scott\inst{14}}
\institute{
Institut d'Astrophysique Spatiale, Bat. 121, 91405 Orsay cedex, France
\and
CNRS 
\and 
Univ. Paris-Sud
\and 
Universit\'e Paris-Saclay
\and
European Southern Observatory, ESO Vitacura, Alonso de Cordova 3107, Vitacura,
Casilla 19001, Santiago, Chile
\and
Atacama Large Millimeter/submillimeter Array, ALMA Santiago Central Offices,
Alonso de Cordova 3107, Vitacura, Casilla 763-0355, Santiago, Chile
\and
Universit\'e de Toulouse, UMS-OMP, IRAP, F-31028 Toulouse cedex 4, France
\and
LERMA, UMR~8112 CNRS, Ecole Normale Sup\'erieure and Observatoire de Paris, Paris, France
\and
Steward Observatory, University of Arizona, Tucson, AZ 85721, USA
\and
Chalmers University of Technology, Onsala Space Observatory, Onsala, Sweden 
\and
Aix Marseille Universit\'e, CNRS, LAM (Laboratoire d'Astrophysique de Marseille) UMR7326, 13388, Marseille, France
\and
CEA-Saclay, F-91191 Gif-sur-Yvette, France
\and
School of Earth and Space Exploration, Arizona State University, Tempe, AZ 85287, USA
\and
Department of Physics and Astronomy, University of British Columbia,
6224 Agricultural Road, Vancouver, British Columbia, 6658, Canada}
\titlerunning{Planck's Dusty Gems: [CII] in the Garnet}
\authorrunning{Nesvadba et al.}  \date{Received / Accepted }

\abstract{We present spatially resolved ALMA [CII] observations of the
  bright (flux density $S_{350}=$400~mJy at 350~$\mu$m),
  gravitationally lensed, starburst galaxy PLCK~G045.1$+$61.1 at
  $z=3.427$, the ``Garnet''. This source is part of our set of
  ``Planck's Dusty GEMS'', discovered with the {\it Planck's} all-sky
  survey. Two emission-line clouds with a relative velocity offset of
  $\sim 600$ km s$^{-1}$ extend towards north-east and
    south-west, respectively, of a small, intensely star-forming
    clump with a star-formation intensity of
  $220~$M$_{\odot}$ yr$^{-1}$ kpc$^{-2}$, akin to maximal starbursts.
  [CII] is also seen in absorption, with a redshift of $+350$ km
  s$^{-1}$ relative to the brightest CO component. 
  [CII] absorption has previously only been found in the Milky
  Way along sightlines toward bright high-mass star-forming regions,
  and this is the first detection in another galaxy. Similar to
  Galactic environments, the [CII] absorption feature is associated
  with [CI] emission, implying that this is diffuse gas shielded from
  the UV radiation of the clump, and likely at large distances from
  the clump. Since absorption can only be seen in front of a continuum
  source, the gas in this structure can definitely be attributed to
  gas flowing towards the clump. The absorber could be part of a
  cosmic filament or merger debris being accreted onto the galaxy. We
  discuss our results also in light of the on-going debate of the origin of
  the [CII] deficit in dusty star-forming galaxies.}

\keywords{galaxies: high-redshift\vspace{-7mm}}
\maketitle
\section{Introduction}
\label{sec:introduction}

The bright 158~$\mu$m line of singly ionized carbon, [CII]158, is one
of the most versatile tracers of the interstellar gas in star-forming
galaxies. With a low ionization potential, C$^+$ is a probe of the
cold neutral gas in galaxies, and can be associated with intensely
star-forming environments \citep[e.g.,][]{stacey10,rigopoulou14}
and diffuse gas \citep[e.g.][]{langer10, gerin15}. It is the main
coolant of the cold neutral medium \citep[][]{bennett94}, and the most
luminous line of gas heated by UV photons from star formation over
wide ranges of density and UV intensity \citep[e.g.,][]{goldsmith12,
  lepetit06, kaufman99}, but can also be bright in shocked gas
\citep[][]{appleton13}. The diversity of environments and gas
conditions probed by [CII]
\citep[][]{rawle13,boone14,schaerer15,knudsen16} entail 
that many empirical properties of [CII] are not yet very
well understood, especially at high redshift.

Here we present spatially resolved ALMA cycle~2 observations of [CII]
in a strongly gravitationally lensed dusty starburst galaxy at
z$=3.427$, G045.1$+$61.1 (``the Garnet''), which was discovered using
th Planck all-sky survey, and was subsequently confirmed with SPIRE
imaging \citep{planck15a}, and through ground-based
observations \citep[][\citetalias{canameras15}
  hereafter]{canameras15}. The Garnet consists of four counter-images
seen behind a small group of galaxies at a spectroscopic redshift of
z$=$0.56 \citepalias[][]{canameras15}, which together reach a peak
flux density of 400~mJy at 350~$\mu$m as seen with SPIRE. The spectral
energy distribution (SED) of the dust is consistent with star
formation, without obvious signs of AGN contamination
\citepalias[][]{canameras15}.  Here we focus on the brightest of the
four counter-images, which contributes 46~\% of the total flux at
850~$\mu$m seen with the SMA \citepalias[][]{canameras15}. The SMA
recovers over 90~\% of the total flux measured with SCUBA-2 at the
same wavelength.  Our ALMA data resolve the [CII] emission into two
clouds at different velocities, around an intensely star-forming
clump, against which [CII] is seen in absorption.  To our knowledge,
no such absorption line has previously been found outside the Milky
Way, where it is seen against massive star-forming clouds and is a
tracer of the diffuse interstellar medium \citep[ISM, ][]{gerin15}.
Throughout the paper we adopt a flat H$_0=$70~km~s$^{-1}$~Mpc$^{-1}$
concordance cosmology with $\Omega_{M}=0.3$ and
$\Omega_{\Lambda}$=0.7.
\vspace{-4mm}

\section{Observations and data reduction}
\label{sec:obs}
\begin{figure*}
\centering
\includegraphics[width=0.9\textwidth]{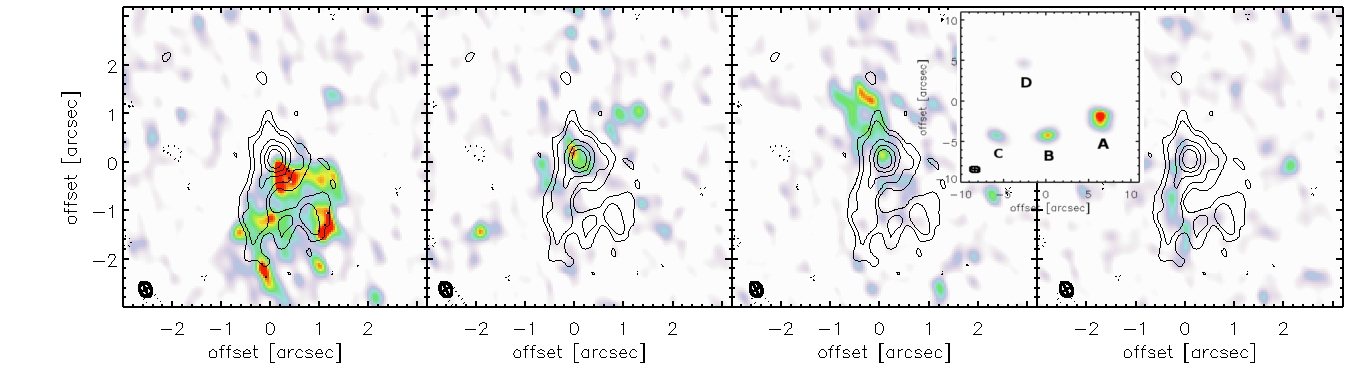}
\caption{\label{fig:intimg} {\it left to right} [CII] channel maps
  centered on $-500$ km s$^{-1}$, systemic velocity, $+100$ km
  s$^{-1}$, and $+350$ km s$^{-1}$, respectively, all relative to
  z$=$3.427 and with a channel width of 87 km s$^{-1}$}. Contours show
  the continuum in the line-free spectral windows and are given
    for 3, 5, 10, 15, and 20$\sigma$. Negative contours are
    $-3\sigma$. The inset shows the SMA 850~$\mu$m image with all
    counter images; the gray circle is the primary beam of ALMA. The other
    panels show image A. The ALMA beam size  in the lower left
    corner of each map corresponds to 150~pc at z=3.4 (for
    $\mu=20$). \vspace{-3mm}
\end{figure*}

\begin{figure}
\centering
\includegraphics[width=0.25\textwidth]{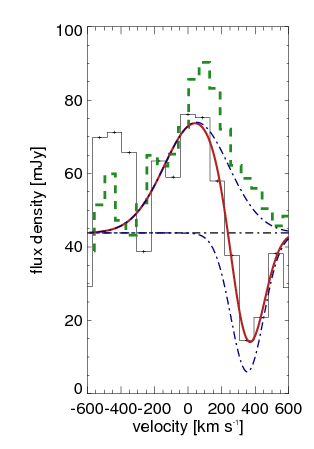}
\caption{\label{fig:spec} Integrated spectrum
in a 1.2\arcsec\ aperture centered on the bright continuum
  source. The spectrum is shown in black with $87$ km s$^{-1}$
  channels, blue lines are individual Gaussian fits, the 
  red line plots the combined fit. The
  continuum (dashed black line) was measured
  independently from the same aperture in the line-free continuum
  spectral windows. The dashed green line shows the integrated [CI]
  1-0 profile.\vspace{-7mm}}
\end{figure}

ALMA observed our Cycle 2 project with 37 antennae on 6 June, 2015 and
obtained 27 min of band 8 data in the C34-3
  configuration, tuned to the [CII] line (428.775 GHz sky frequency)
in one spectral window. Three more windows were placed 
on the line-free
continuum. The intrinsic channel width was 15.6 MHz in all spectral
windows, yielding a total bandwidth of 1.856 GHz per window
after flagging the edge channels. This corresponds to about 1300 km
s$^{-1}$ at z=3.4, sufficient to cover all [CII] emission, assuming a
line profile similar to CO(4-3), which we used as reference. One
spectral window had four times higher noise levels on average, which is due to
the nearby telluric oxygen line at 424.8~GHz and was ultimately not used
in the analysis.

We used the standard manual scripts for ALMA reduction with the Common
Astronomy Software Application {\tt CASA}, and {\tt CLEAN} to
construct the synthesized beam-deconvolved images of the frequency
data cubes.  The spectral channels were rebinned to a width of
50~km~s$^{-1}$. The major and minor axis size of the beam is
0.41\arcsec$\times$0.27\arcsec\ along PA$=$13.5$^\circ$, The largest
angular scale is 3.9\arcsec, and the primary beam size 12\arcsec. The
rms is 1.3~mJy in each 50~km~s$^{-1}$ channel, and 0.46~$\mu$Jy in the
averaged continuum image extracted from spectral windows~1 and 2.

In addition, we also used the CO(4-3) line from
\citetalias{canameras15} and the [CI] line corresponding to the
$^3P_1-^3P_0$ transition of atomic carbon at 492.16~GHz in the rest
frame, which we observed during the same campaign with EMIR
 on the IRAM 30 m telescope with a 20\arcsec\ beam. 
Our [CI]
detection also confirmed the redshift of the Garnet. At
$z=3.4266\pm0.0005$, the line falls at
(111.183$\pm$0.015)~GHz. Observations were carried out on 2 March,
2014, under good and stable conditions, for a total observing time of
81~min. Observations and the data
reduction are described in \citetalias{canameras15}. The integrated
apparent line flux of [CI](1-0) is (8.6$\pm$1.6)~Jy~km s$^{-1}$, with
a peak brightness temperature of 3.3~mK, and
FWHM$=(420\pm$120)~km~s$^{-1}$. The integrated
  [CI] luminosity is $\mu L^\prime=(3.3\pm0.7)\times 10^{11}$ K km
  s$^{-1}$ pc$^2$, and we estimate a gas mass of $(17.4\pm4)\times
  10^{11} \mu^{-1}$ M$_{\odot}$, a factor of~2 higher than
previously estimated in \citetalias{canameras15} from CO(4-3), and
broadly confirming our previous choice of a low CO-to-H$_2$ conversion
factor in \citetalias{canameras15}. All values are uncorrected for the
gravitatonal magnification factor $\mu$. The $^3$P$_2-^3$P$_1$
[CI](2-1) line at 809~GHz is inaccessible from the ground at z=3.43.
We made this estimate using the relation reported in \citet{papadopoulos04}
and \citet{wagg06}, $M^{{\rm [CI]}}({\rm
  H_2})=1375.8\frac{D^2}{1+z}\ X_{{\rm [CI,-5]^{-1}}}\ A_{10,7}^{-1}
Q_{10}^{-1} S_{{\rm [CI]}} [{\rm M}_{\odot}]$, where $D^2$ is the
luminosity distance at redshift z, $X_{{\rm [CI,-5]}}=5$ the abundance
of atomic carbon in the ISM in units of $10^{-5}$, and $A_{10}$ the
Einstein coefficient in units of $10^{-7}$ s$^{-1}$. $Q_{10}=0.7$
captures the population of the upper and lower level of the transition
\citep[][]{papadopoulos04} \vspace{-5mm}

\section{Observed morphologies and spectrum\vspace{-3mm}}
\label{sec:morphologies}

We used the line-free continuum image from spectral windows 1 and 2 to
estimate the continuum flux density in each spaxel of the [CII] cube
and subtracted this value from every spectral bin. We also corrected
for the (insignificant) brightening of the dust continuum by
2\%\ between the line and continuum windows expected for an
apparent dust temperature, $T_d=36$~K \citepalias{canameras15}.

Fig.~\ref{fig:intimg} shows the channel maps of the two extended
emission line regions (EELRs) as seen in the image plane, which
  are offset by 564~km s$^{-1}$ from each other (redshifted and
blueshifted channels, respectively), and centered on a small
continuum clump with high surface brightness. The integrated spectral
properties of each region are listed in Table~\ref{tab:lines}.  Each
EELR extends about $\sim2$\arcsec\ radially from the continuum source,
but the southwestern region is much wider in tangential
direction, 2.2\arcsec\ instead of 0.9\arcsec\ measured for the northeastern
region (all sizes are 3$\sigma$ isophotal sizes).This might either
represent differences in the intrinsic gas properties or in the
gravitational magnification.  In either case, the strong sudden
velocity offset, which is larger than the FWHM of the individual
clouds, shows that the [CII] lines do not probe large-scale rotational
motion, but are two kinematically separated clouds.

\begin{table*}
\centering
\begin{tabular}{lccccccc}
\hline
Component & $\mu\ I_{\rm CII}$ & Velocity & FWHM & $\mu\ L^\prime_{\rm CII}$ & $\mu L_{\rm CII}$ & $\mu$ \\
          &[Jy km s$^{-1}$] & [km s$^{-1}$] & [km s$^{-1}$] & [$10^{10}$ K km s$^{-1}$ pc$^2$] & [$10^9 L_{\odot}$] & \\ 
\hline
Blue EELR &  52.4$\pm$0.6 & -467$\pm$15 & 202$\pm$20 & 9.5$\pm$0.7  & 21$\pm$0.2  & 21\\
Red EELR  &  8.3$\pm$0.6  & 97$\pm$36   & 213$\pm$50 & 1.5$\pm$0.2  & 3.3$\pm$0.2 & 22 \\
Clump     &  4.8$\pm$0.6  & -418$\pm$47 & 190$\pm$66 & 0.87$\pm$0.1 & 1.9$\pm$0.2 & 10  \\
          &  8.73$\pm$0.8 & 0$\pm$35    & 274$\pm$49 & 1.5$\pm$0.1  & 3.5$\pm$0.3 & 10  \\
\hline
Absorber  & -0.60$\pm$0.15  & 343$\pm$34& 124$\pm$25 & $\cdots$ & $\cdots$ & 10\\
\hline
\end{tabular}
\caption{\label{tab:lines} Emission-line properties for the redshifted
  and blueshifted EELRs and the two line components seen against the
  clump. For the absorption line, we list the depth $\tau$ of the
  flux-normalized spectrum instead of the flux.\vspace{-7mm}}
\end{table*}

We used {\tt Lenstool} \citep[][]{kneib96,jullo07} to construct a
gravitational lens model from the positions of the four images seen in
Fig.~2 of \citetalias{canameras15}, and calculated the magnification
of each image, assuming that all four are multiple images of the same
region. The average magnification at the position of the bright
star-forming clump is $\mu=10$, while $\mu=21$ and $\mu=22$ for the
blueshifted and redshifted gas, respectively. We will present a more
detailed analysis of the lensing configuration when our scheduled
Hubble Space Telescope imaging has been taken (GO-14223/PI~Frye). The
present analysis does not strongly depend on the details of the lens
modeling, because the most important physical parameters (surface
brightnesses, velocities) are not directly affected by the lensing.

The continuum clump is marginally resolved, with a FWHM size of
0.66$\pm$0.01\arcsec$\times$0.51$\pm$0.01\arcsec\ along the major and
minor axis, respectively, compared to a beam size of
FWHM$=$0.41$\times$0.27\arcsec.  The continuum flux density of the
clump within 2$\times$ the FWHM beam size is 42.9~$\mu^{-1}$ mJy,
corresponding to that expected for a modified blackbody with
$\beta=2.0$ and $T=36~K$\citepalias[][]{canameras15}. The clump alone
contributes 28\% of the total FIR luminosity of G045.1$+$61.1. Another
18\% comes from faint, diffuse emission associated with the two
emission-line regions (Fig.~\ref{fig:intimg}), the remaining flux
comes from the other lensed images shown in the inset of
Fig.~\ref{fig:intimg}. Adopting a conversion from $L_{\rm FIR}$,
integrated between 8~$\mu$m and 1000~$\mu$m, and star formation rate,
SFR, of SFR\ [M$_{\odot}$ yr$^{-1}$]$=4.5\times\ 10^{-37}$~W
\citep[][]{kennicutt89}, and the apparent size of the clump of
0.66\arcsec$\times$0.51\arcsec, this implies an average star-formation
intensity of 220 M$_{\odot}$ yr$^{-1}$ kpc$^{-2}$, which is in the range of
maximal starbursts \citep[][]{elmegreen99}.

In Fig.~\ref{fig:intimg} we also show the spectrum extracted from an
aperture with 1.2\arcsec\ diameter centered on the bright continuum
source. As expected from the strong velocity jump in this region
(also visible in the channel maps), we see both velocity components,
one at $+100$~km s$^{-1}$ from the systemic redshift of $z=3.427$
measured at the [CI] peak and one at $-500$ km s$^{-1}$. Line emission
is faint near the bright star-forming clump; most flux comes from the
extended blueshifted and redshifted EELRs (Fig.~\ref{fig:intimg} and
Table~\ref{tab:lines}).  In addition, we also see an absorption
feature around velocities of $+$350 km s$^{-1}$. The depth of the
absorption trough was measured from the normalized spectrum to be
$-0.60\pm 0.15$, and other fit parameters are listed in
Table~\ref{tab:lines}.

We examined whether this feature might be an
observational artifact. The line is resolved into at least five spectral
channels and does not depend on the specific choice of aperture size
or \textit{CLEAN} parameters used in the reduction. It is also
apparent in the dirty maps. We do not detect any strongly negative
regions in the maps, and our field does not contain bright sources
that could cause strong negative signals. The spectra in the
continuum spectral windows are flat, and neither the check source
nor the phase calibrator show a corresponding amplitude drop. We
therefore conclude that there is no evidence that the
absorption-line feature in G045.1$+$61.1 is spurious.\vspace{-5mm}

\section{Astrophysical nature of [CII] in the Garnet}
\label{sec:nature}

Our ALMA data show a rich environment with two kinematically offset
emission-line clouds around a bright star-forming clump, and another
component is seen in absorption. For a magnification factor of
$\mu\sim10$, the FIR luminosity of the clump implies 
$\sim$400~M$_{\odot}$ yr$^{-1}$ of star formation, an order of magnitude
greater than in massive star clusters in the Milky Way and nearby
galaxies \citep[$0.1-{\rm few} \times 10$ M$_{\odot}$
  yr$^{-1}$,][]{larsen00}.

We used the PDR models of \citet{kaufman99} to derive the average gas
properties from the [CII] luminosities obtained with ALMA, the CO(4-3)
luminosity of \citetalias[][]{canameras15}, and the [CI] 1-0
luminosity (Sect.~\ref{sec:obs}). From the ratios $L_{{\rm CII}}/L_{{\rm
    CI}}=72.5\pm$13 and $L_{{\rm CII}}/L_{\rm CO}=23\pm$1, we estimate
that the gas in the Garnet is on average exposed to a radiation field
$200$ times greater than in the solar neighborhood, and has a density
of about $10^4$ cm$^{-3}$. To derive these line ratios, we corrected
the [CI] 1-0 and CO(4-3) luminosities by a factor of 0.46 to take into
account that the fainter images are also in the beam, and we assumed that
the gas probed by the CO line emission is optically thick.

[CII] emission extends over at least 1200 km s$^{-1}$ in the Garnet,
over a small area of $\lesssim 1$~kpc in the source plane
(assuming $\mu\ge 10$ as suggested by our preliminary lensing
  analysis, see Sect.~\ref{sec:morphologies}). It is therefore
interesting that the Garnet has only redshifted, and no blueshifted
[CII] absorption. Unlike EELRs, where blue- and redshifts may either
indicate outflows or inflows from the host galaxy (because the gas may
be in front of the galaxy or behind it), redshifted absorption is
always found in front of the emitter, and therefore a unique signature
of an inflow, not an outflow. The [CI] emission line shows a
distinct wing at the same velocity as the [CII] absorption. This is
also characteristic of [CII] absorption in the Milky Way. 
\citet{gerin15} have shown that both lines can simultaneously arise
from gas with similar conditions. By analogy with the Milky Way, we
also assume that the line we see consists of multiple deep
narrower absorption components that sample a velocity range of 120 km
s$^{-1}$. This might be the intrinsic velocity range of multiple
clouds in a filament or merger debris. Even in the Milky Way, where
average velocity dispersions in the ISM are about 10 km s$^{-1}$,
the total width over which absorption is found is known to be up to $70-80$ km
s$^{-1}$.

We followed \citet{gerin15} to estimate a column density of the
absorbing gas. With their Eq. (1) originally taken from
\citet{goldsmith12} to derive a C$^+$ column density, $N(C^+)=1.4\times
10^{17} \int{\tau dv} [\rm {cm}^{-2}]$, we find $N(C^+)=8.2\times
10^{18}$ cm$^{-2}$, corresponding to $N({\rm HI})=5.9\times 10^{22}$
cm$^{-2}$ when adopting  the Galactic carbon abundance of
$1.4\times 10^{-4}$. This choice is reasonable for
massive dusty starburst galaxies at high redshift, which have gas-phase 
metallicities comparable to those in the Milky
Way \citep[][]{takata06,nesvadba07}, but might be too high for
infall of more pristine gas. For
solar metallicity, this column density is about a factor of~2 higher
than the most extreme values found in the Milky Way, and is accordingly
more suitable for lower metallicity gas. The implied total column density of
neutral gas is comparable to those estimated for H$_2$ from the CO
emission-line surface brightness \citep[][Canameras et al., 2016, in
  prep.]{swinbank11} and also plausible here, given our gas mass
estimate from [CI].

It is impossible to estimate a total mass from an absorption line,
therefore we constrained the total mass within this region from the
flux in the red wing of [CI], that coincides in velocity with the
[CII] absorption (Fig.~\ref{fig:intimg}). The wing contains 26\%\ of
the [CI] 1-0 flux, and $3.2 \times 10^{10}$ M$_{\odot}$ in mass (for
$\mu\sim10$), comparable to the total molecular gas content of high-z
starbursts \citep[e.g.,][]{tacconi08}. Carbon is easily ionized
\citep[e.g.,][]{goldsmith12}, and the absorber must therefore be
effectively shielded from the intense UV radiation of the clump, and
cannot be associated with intense star formation. The absorber is most
likely located at a large distance from the clump, potentially several
kpc. A velocity offset of 350 km s$^{-1}$ is also too large for the
gas to be associated with the clump itself, and is comparable to
velocity dispersions of massive low-redshift galaxies
\citep[][]{bernardi05}, suggesting the gas is bound to the host
galaxy. The Garnet provides a rare opportunity to study gas in a
high-redshift galaxy outside of intense star-forming regions, and
possibly even infalling gas from a cosmic filament, merger debris or
satellite accretion.

The blue- and redshifted EELRs extend from a massive
star-forming clump, which could be probing a wind. We used the recent
empirical analysis of \citet{heckman15} of momentum-driven winds in
intensely star-forming low-redshift galaxies to show that this
scenario is unlikely, and alternative explanations, such as overlapping
gas clouds in a star-forming region or merger, are more
realistic. \citet{heckman15} argued that winds will escape if the
momentum input from star formation is $\ge10\times$ a critical value
that depends on the column density, $N_C$, and circular velocity,
$v_c$, of the galaxy. We used the 400~km s$^{-1}$ velocity of the
absorbing gas to approximate the circular velocity of the Garnet at
large distances.

Using Eq.~(3) of \citet{heckman15}, we found
$p_{crit}=2\times 10^{36}$~dyn, for a fiducial cloud distance of
100~pc from the starburst. The critical momentum increases linearly
with increasing distance. We also adopted $N_{\rm C}=1\times 10^{23}$
cm$^{-2}$, $v_{\rm c}=$400~km~s$^{-1}$, and used the mass per $H_2$
molecule, that is $3.24\times 10^{-24}$ g. $N_{\rm C}$ is set by 
the gas mass surface density estimates of 0.4 and $3.9\times 10^9$
M$_{\odot}$ kpc$^{-2}$ implied by the [CI] line profile, assuming
that [CI] and [CII] have the same morphology. Assuming \citep[again
  following][]{heckman15} that the combined momentum flux from ram
and radiation pressure in the star-forming clump is $4.8\times
10^{33}$~SFR dyn, we find a total momentum input from star formation
of $1.9\times 10^{36}$ dyn for SFR$=$400~M$_{\odot}$ yr$^{-1}$.
  Following \citet{heckman15}, this may be enough to balance
  gravity, but not to form a wind that escapes, the main difference to
  low-redshift galaxies being the high column densities. Of course,
  this does not rule out the presence of fainter more energetic wind
  components, as reported for unlensed galaxies
  \citep[][]{maiolino12,cicone15}.
\vspace{-6mm}

\section{Implications for high-z galaxies and conclusions}
\begin{figure}
\centering
\includegraphics[width=0.45\textwidth]{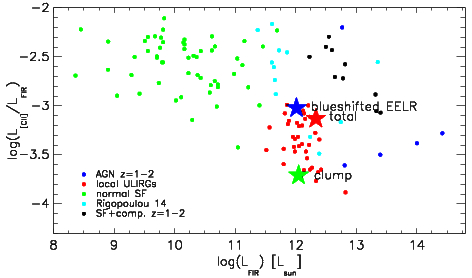}
\caption{Ratio of [CII] to FIR luminosity as a function of FIR
  luminosity for the Garnet and for several samples of low and
  high-redshift galaxies in the literature. The FIR luminosities have
  been corrected for average magnification factors of 10, 21, and 21
  at the position of the total source, and the blueshifted gas and the
  continuum source, respectively.\label{fig:fircii}\vspace{-2mm}}
\end{figure}

It is also interesting to investigate how our sources relate to dusty
star-forming galaxies at high redshifts in the field, where we have
more comprehensive knowledge of the global properties, but lack the
detailed spatially resolved information that only strong lensing can
provide. Infrared-selected, intensely star-forming galaxies at low
redshift show a marked deficit in [CII] luminosity relative to the
total FIR luminosity, compared to galaxies with lower star formation
rates and less extinction; this is a trend that is rarely found
at high redshift \citep[][Malhotra et al. 2016]{rigopoulou14}.  We
show in Fig.~\ref{fig:fircii} where the Garnet lies relative to other
populations of low- and high-redshift galaxies. The
absorption does not lower the continuum luminosity, which is derived
from the integrated flux between 8 and 1000$\mu$m in the
rest-frame. The blue EELR falls into the region of the diagram, in which
distant and nearby star-forming galaxies, and local ULIRGs overlap,
whereas the aperture centered on the clump lies near the extreme end
of the local ULIRG sample. This highlights the fact that unresolved
observations of high-redshift galaxies sample wide ranges of intrinsic
line ratios. The integrated spectrum of the Garnet lies very close to the
blue EELR, suggesting that the global [CII]/FIR ratios are dominated
by the global spatial distribution and sizes of gas clouds within
star-forming regions, and do not reflect the properties of individual
star clusters.

These ALMA data of G045.1$+$61.1 at $z=3.43$, the first of
Planck's Dusty GEMS, shows that [CII] at high redshift probes a wide
range of environments, from the dense gas irradiated by UV photons
from maximally star-forming clumps to the diffuse interstellar gas at
large distances seen in absorption. Although we found signatures of
[CII] spanning at least 1200~km~s$^{-1}$, we did not detect strong
evidence for a wind that would regulate star formation by removing the
gas reservoirs, as is often suggested in starburst-driven wind
scenarios. This might indicate that broad wings of [CII] and other
lines in high-redshift galaxies probe a much richer
phenomenology than previously thought. Deeper higher-resolution
observations spanning a wider spectral range will open an interesting
new window to study the cold neutral gas in intense high-redshift
starburst galaxies, including the ambient gas outside of the intense
star-forming regions. \vspace{-3mm}

\section*{Acknowledgments}
We thank the ALMA staff for carrying out the observations, and the
team at the ARC in Grenoble for their help with preparing the data
taking. We also thank the referee for suggestions that helped
improve the paper. NPHN wishes to acknowledge funding through the JAO
within their visiting scientists program, and is grateful for their
hospitality during her stay as a science visitor in
Vitacura.{\vspace{-3mm}}

\bibliographystyle{aa}
\bibliography{lens}

\end{document}